# A Note on Several Meteorological Topics Related to Polar Regions

Krzysztof Sienicki

*Chair of Theoretical Physics of Naturally Intelligent Systems, Topolowa 19, 05-807 Podkowa Leśna, Poland*

krissienicki@yahoo.com

(24 July, 2011)

## Abstract

Analysis of the meteorology of Polar Regions is fundamental to the process of understanding the global climatology of the Earth and Earth-like planets. The nature of air circulation in a polar vortex is of preliminary importance. I have show that the local and continental spatiotemporal relationship between near surface wind events is self-organized criticality. In particular, the wind event size, wind event duration, and duration of quiescent wind event are well approximated by power-law distributions. On a continental scale, the wind events in the Antarctic tend to be self-organized criticality with ergodic properties. A similar self-organized criticality wind event was also found in Taylor Valley located at McMurdo Dry Valleys discovered by Captain Scott's expedition. Captain Scott's meteorological *Terra Nova* record was also examined. I have also revisited and re-analyzed wind events in Hornsund at Spitsbergen Island, in terms of marginal probabilities and marginal copulas which describe positive Lévy process.

# Contents

## 1. Introduction

Spatiotemporal relationships between processes are of fundamental importance in physics and meteorology. For example, in the case of special theory of relativity the connection between processes in inertial frames of reference for homogeneous space and time is described by the Lorentz transformation. However, the transition between non-inertial frames or classical-quantum limit is less obvious and understandable.

Statistical mechanics is a unique combination of the law of large numbers and the laws of mechanics. Since the development of statistical mechanics in the late 19$^{th}$ century, the theory was successfully applied in description of many equilibrium and non-equilibrium physical systems. The central limit theorem and the law of large numbers characterize the macroscopic statistical properties of large ensembles of independent and identically distributed random variables. These two probability laws give rise to ensembles' aggregate statistics described by either Gaussian or Lévy probability formulas.[1]

Although we are all familiar with Gaussian law, it may be noted that this law does not result in the obvious way from the above mentioned probability laws. Provided the mean of a random variable is zero and that variable variance is finite, the Gaussian law is rather the limit of large ensembles of independent and identically distributed random variables. In the case of extreme values (maximum, minimum), the central limit theorem leads to the extreme value probability theory governed by the Fréchet, Weibull, or Gumbel distribution functions.[2]

Use of mathematical formula for Lévy distribution was recommended long ago. It has only been during the past decade, however, that observations have been reported of many physical systems not obeying the law of large numbers. These systems are characterized by dominating large and rare fluctuations which give rise to broad distributions with power-law tails. Both Gaussian and Lévy distributions are stable distributions – a linear combination of their random variables leads to Gaussian and Lévy distributions, respectively.

In physical terms, random walks can be classified either as Brownian or Lévy random walks. In the first case, the step lengths $\ell_i$ have a characteristic scale defined by the first and second moment (mean and variance, respectively) of the step length density distribution $\mathcal{P}(\ell)$. In the second case, the Lévy random walk, the walk step lengths have no characteristic scale. The second moment (variance) diverges, and in some cases even the first moment; the mean, diverges. The Lévy distribution has self-affine properties: $\mathcal{P}(\lambda\ell) \sim \lambda^{-\alpha}\mathcal{P}(\ell)$ for $1 < \alpha \leq 3$.

One of the most important differences between Gaussian and Lévy distributions is that the former one admits arbitrary (infinite) variance and therefore the possibility of large deviations. This particular property of the Lévy distribution was used in studies of Lévy flights and search processes.

There is hardly anything more fundamental for understanding statistical mechanical systems than whether system it is ergodic or not. The nation or ergodic hypothesis was present in early Boltzmann writings on the kinetic theory of gasses.[3] According to this hypothesis, if the system is left in free evolution and waiting for a sufficiently long time, the system will pass through all the states consistent with its general condition.[4] Ergodic hypothesis connected the phase average of a function, with a infinite time average of a physical quantity of the system. However, it was proven that a trajectory obeying the ergodic hypothesis passing through any phase point, cannot be a mechanical trajectory.[5] For this reason, a new notion was developed with equality of phase averages and infinite time averages using only dynamical properties of the system under consideration. The ensemble average and time average of observables are equal in the infinite-time limit.

In this article, I want to investigate how the wind events are organized from a dynamical and statistical point of view. In particular I will analyze wind events in the Antarctic and the possibility that they are ergodic self-organized criticality.

## 2. Self-organized Criticality

In the mid eighties, the notion of self-organized criticality was put forward. In recent years, this notion was developed and explored in many fields of science and human activities.[6] All studied dynamical systems are characterized by the possession of a critical point as an attractor. Their macroscopic behaviour shows spatiotemporal scale invariance which is characteristic to a given critical point of the system.

Power law distributions usually occur in one of two forms. In the first, the probability of an event of magnitude is expressed as a power of the magnitude itself. In the second, the common form of power law distributions are rank ordered. Rank 1 has the most frequent or probable event or state; rank 2 has the second most probable, *ect*.

A real valued random variable $X$ with cumulative distribution function $F(x) = \mathbb{P}\{X \leq x\}, x \in \mathbb{R}$ is said to have (right, $x \to \infty$) heavy tail if

$$\mathbb{P}\{X > x\} = 1 - F(x) = \mathcal{L}(x)x^{-\alpha}, as\ x \to \infty \quad (1)$$

for some $\alpha > 0$, where $\mathcal{L}(x) > 0$ is a slowly varying function, an asymptotically small function, $\lim_{x \to \infty} \mathcal{L}(bx)/\mathcal{L}(x) = 0$ and constant $b > 0$. The so-called tail *exponent* or *scaling* parameter $\alpha > 0$ controls the rate of decay of $F(x)$ and hence characterizes its tail behavior.

Equation (1) can be written in terms of probability distribution with density

$$\mathbb{P}\{x \leq X \leq x + dx\} = ax^{-\alpha}dx \quad (2)$$

Let $p(x)$ be the power-law density given by the above equation and $p(x) = ax^{-\alpha}$. The scaling of $x$ by a constant factor $c$ leads to only a proportionate scaling of the power-law density itself, i.e. $p(cx) = a(cx)^{-\alpha} = ac^{-\alpha} \cdot x^{-\alpha} \propto p(x)$. In theory, the scale invariance is a fundamental characteristic of objects or laws that do not change under the change of scales. However, in real physical and other applications, the scale invariance is somehow limited, though it may cover many decades.

Equation (2) is continuous power-law probability distribution density which diverges for $x \to 0$. Therefore, the lower bound $x_{min} > 0$ to the power-law density $p(x)$ is introduced, and for discrete values takes the form

$$p(x) = \frac{x^{-\alpha}}{\varsigma(\alpha, x_{min})} \quad (3)$$

where $\varsigma(\alpha, x_{min})$ is Hurwitz zeta function, $\varsigma(\alpha, x_{min}) = \sum_{n=1}^{\infty}(x_{min} + n)^{-\alpha}$.

## 3. Vortex over Antarctic Continent

The Antarctic continent is the highest, driest and coldest of the continents on the Earth. About 14 million square kilometers of Antarctic land area is essentially covered with ice and snow. A tiny 0.32 % of exposed land at the coastal fringes is snow and ice free. The entire size of Antarctic area doubles (in September) during the astral winter. Unexposed land or the Antarctic Archipelago is covered by ice which forms the Antarctic dome. This area rapidly rises from the coast to reach about 4000 m. The area of the entire Antarctic dome cross section taken at a height of 3000 m is the size of Australia.[7]

When investigations were first made, many authors assumed that katabatic winds are downslope gradient-driven flows.[8] The long-wave radiative loss to space, leads to the build-up of high density cold air over the Antarctic dome. The presence of dome slope (relative height) induces a horizontal temperature gradient, leading to a downslope horizontal pressure-gradient katabatic force. Katabatic winds are then driven toward the coastal fringes and outwards towards the Antarctic Ocean. This scenario may be pertinent during the winter time. However, during the equinoctial seasons, katabatic winds are also observed. Such an observance takes place despite the simple fact that radiative loss, mentioned above, does not takes place. On the contrary, the Antarctic dome surface gains energy. While using loosely defined parameters, various rationalizations have been developed. These rationalizations account for the expected different dynamical behaviour at *meso*- to synoptic-scale regimes of driving forces of near surface wind in the Antarctic.[9]

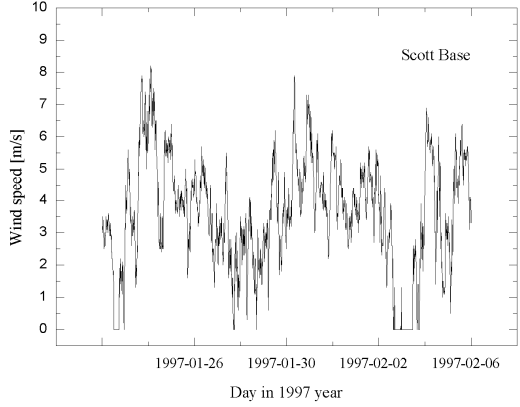

Figure 1. An example of a near surface wind event (wind speed $v(t)$) measured at the Antarctica New Zealand Scott Base.

However, even a cursory examination will show that the vortex over the Antarctic continent is slowly-driven by a solar radiation non-equilibrium system with many degrees of freedom and a high level of nonlinearity. Figure 1 depicts a near surface wind speed $v(t)$ measured at the Antarctica New Zealand Scott Base station located at the southern tip of Hut Point Peninsula of Ross Island, Antarctica. The erratic behavior of this time series is clearly visible.

In equilibrium statistical mechanics, the inverse of temperature $(1/k_B T)$ is a constant physical system parameter for a given state with energy $E$. However, all natural phenomena exhibit significant spatiotemporal temperature fluctuations (*meso*scale). For this reason, such phenomena must be regarded as non-equilibrium systems. In order to illustrate this behaviour, I will analyze the event-like structure of wind events partially depicted on Fig. 1. Their shown time scale can be divided into two compartmental events directly related to the size and the duration of the events. The first case of non-zero measurements of the wind speed I will call the wind event size($w_s$). I define the wind event as $w_s = \int v(t) dt \approx \sum v(t) \Delta t$ for *successive* non-zero wind speeds, where $\Delta t$ is the size of the measurement bin. In the second case, the second compartment can be considered duration of wind event ($w_t$) and duration of quiescent wind event ($w_q$), respectively.

To analyze wind behaviour, I will use its size distribution and occurrence frequency distribution. When estimating a cumulative *size* distribution $p_s(w_s)$, wind *duration* distribution $p_t(w_t)$ and wind *quiescent* distribution $p_q(w_q)$ for a given data set, the number of wind events with size $w_{s,t,q} \geq 0$ or greater is counted.

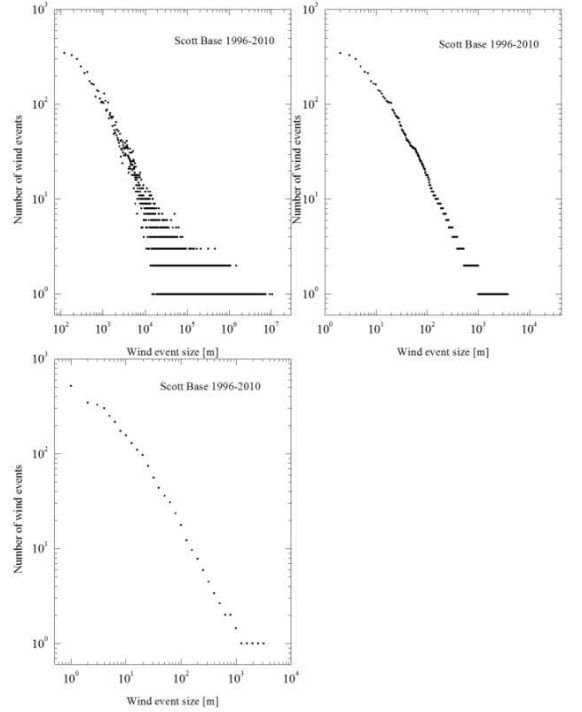

Figure 2. The cumulative size distribution function $p_s(w_s)$ for "raw" data and ranked and binned data for the Antarctic New Zealand Scott Base station.

On Fig. 2 I depicted the cumulative size distribution function $p_s(w_s)$ for "raw" data and ranked and binned data ($b = \log(x_{i+1}) - \log(x_i)$). Fading of fat tail due to the binning procedure is self-evident. On the same figure, a power-law fit of $p_s(w_s) \propto w_s^{-\alpha_s}$ (solid line) is also shown. The best fit was obtained for scaling parameters of wind event size $\alpha_s = -1.13 \pm 0.01$.

For the remaining wind events I used similar cumulative size distributions

$$p_t(w_t) \propto w_t^{-\alpha_t} \quad and \quad p_q(w_q) \propto w_q^{-\alpha_q} \qquad (4)$$

and ranking, binning, and fitting procedures, with the result: duration of wind event $\alpha_t = -1.39 \pm 0.03$ and quiescent wind event $\alpha_q = -2.68 \pm 0.06$, respectively.

Interestingly, all three probability distributions investigated at the Scott Base station follow a power-law. Moreover, probability distributions depicted on Fig. 2 were calculated for the entire wind record at the Scott Base, without season or day time differencing. The uniformity of these probability distributions suggests that there is some fundamental phenomenon which drives the wind events at the Scott Base station. Macroscopic behaviour of wind, displays the temporal scale invariance governed by power-law distributions.

Currently, there are about 150 weather stations scattered across the Antarctic continent.[10] One third of the stations are manned stations while the remaining ones are automated weather stations.[11] I have performed respective calculations of scaling parameters $\alpha_{s,t,q}$ for more than 40 stations scattered in the interior and costal fringes of the Antarctic with the following extreme (boundary) results:

$$\alpha_s = \langle min, max \rangle = \langle 0.88 \pm 0.03, 1.41 \pm 0.01 \rangle$$
$$= \langle Amundsen - Scott, Pegasus\ S \rangle$$
$$\alpha_t = \langle min, max \rangle = \langle 0.98 \pm 0.05, 1.72 \pm 0.04 \rangle$$
$$= \langle Halley, Pegasus\ S \rangle$$
$$\alpha_q = \langle min, max \rangle = \langle 1.09 \pm 0.09, 4.69 \pm 0.08 \rangle$$
$$= \langle Vito, Explorer \rangle$$

Calculated values of scaling parameters show spatial dependence and vary across the Antarctic continent. At some locations, for example the Amundsen-Scott South Pole station, I could not obtain quality fit to wind data. This inability was due to limited data set as well as due to apparent random and uncorrelated wind speed fluctuations at this site. Similar wind behaviour was also observed at Concordia station.

In spite of an inability to find reasonable fit to wind data from few of the weather stations, the remaining wind data were very well described by power-law distribution. This demonstrates that the wind at the majority of these locations (stations) across the Antarctic continent is self-organized criticality. Thus, I conclude that the wind over Antarctica is self-organized criticality.

## 4. Ergodicity of Wind Events

In the preceding section, I have shown that in the Antarctic their exists many coexisting self-organized wind attractors with scaling parameters independent of initial local (station) conditions. This leads me to picture the critical state of wind event as the union of many coexisting naturally stable attractors. At the same time, there is clear geographical location dependence of scaling parameters $\alpha_{s,t,q}$. However, this geographical location dependence should not be understood as a simple geographical coordinate dependence but rather as interplay between local orography and wind relief.

A long time ago Boltzmann formulated a hypothesis, later know as the *ergodic hypothesis*. According to this hypothesis, leaving a system in free evolution and waiting for a sufficiently long time, will mean the system will pass through all the states consistent with its general conditions, namely with the given value of the total energy. The ergodic hypothesis is a fundamental issue in statistical physics.

The original formulation of the hypothesis was a subject of vigorous research and debate. Ergodicity of a physical system would guarantee that a Gibbs' microcanonical ensemble could be relied on to produce correct results.

A fundamental contribution in the field was provided by Khinchin[12] who assumed that the ergodicity of a physical process should be related to irreversibility of the physical process. The Khinchin theorem tells us that if the autocorrelation function vanishes at infinity, then the process is ergodic. Changes of autocorrelation function represent a measure of how well events from the past predict future events. In Khinchin theorem, there is *a priori* assumption that the second central moment (variance $Var(X)$ and therefore autocorrelation) of the physical process under consideration is finite

$$Var(X) = \int (x - \langle x \rangle)^2 p(x)dx < \infty. \quad (5)$$

Certainly this assumption of finite variance is fulfilled by the great number of physical processes which are governed by the Gaussian distribution function. However, there is a large family of physical systems which do not obey the law of large numbers and the Gaussian distribution function. The behaviour of these systems is a subject of rare and very rare burst-like fluctuations[13] with power-law tails. The feature of these distribution functions, originally developed as a mathematical theory by Lévy, is the divergence of their first and/or second moment. In the case of the above discussed power-law distribution, for $\alpha_{s,t,q} < 2$ gives

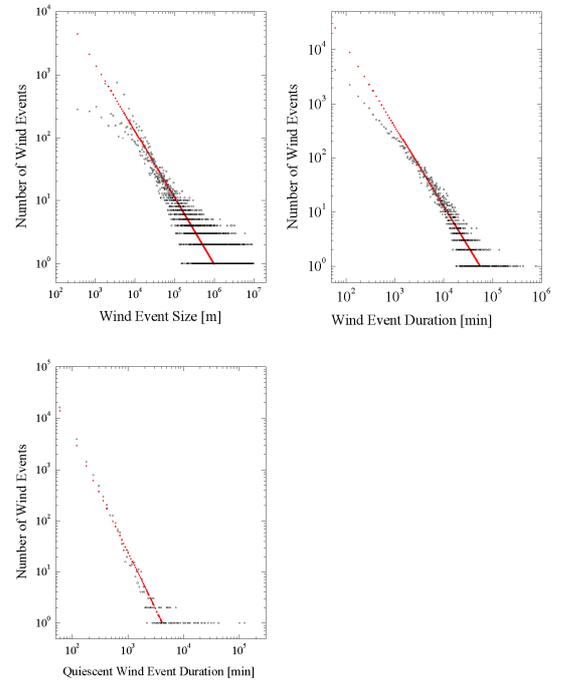

Figure 3. Combined wind event size, wind duration, and wind quiescent event data measured at about 40 stations/locations scattered across the Antarctic continent.

$$\langle x \rangle = \int_0^\infty x p(x)dx \approx \int_{x_{min}}^\infty x p(x)dx$$
$$= \frac{a}{2-\alpha}[x^{-\alpha+2}]_{x_{min}}^\infty \to \infty \quad (6)$$

The mean (average) of power-law probability function *diverges* for scaling parameter $\alpha_{s,t,q} < 2$. Lévy-flight-type distribution functions were recently reported in many systems ranging from animal species foraging patterns[14] to econophysics.

Returning to the analysis of wind events in the Antarctic, it should be noted that all values of scaling parameters

$\alpha_{s,t} < 2$. This clearly indicates that the mean $\alpha \in ]0,2]$ value of wind event size and wind event duration are *non-existent* as the above integral diverges to infinity.

The formal averaging of the near surface wind record in the Antarctic, although technically possible, is not a neither mathematically nor physically justifiable procedure. While having a finite number of wind measurements, finite mean can be calculated. However, to calculate true mean, one has to calculate the infinite number of wind data, without which calculation of mean of power-law distributed data may lead to an entirely wrong and incorrect result.

In the case of random variables with finite variance, as for example with Gaussian distribution, the central limit theorem tells us the sum of random variables will tend to a Gaussian distribution. Later, I will discuss a similar theorem for marginals as well as to copulas related to weather.

Recently Weron and Magdziarz[15] derived a version of the Khinchin theorem for Lévy flights. Their result gives us a tool to study ergodic properties of Lévy $\alpha$ − stable distributions, including the above reported power-law distributions of wind events in the Antarctic.

One should expect that the sum of a number of random variables with power-law tail distributions decreasing as $x^{-\alpha}$ will tend to a stable distribution as the number of variables grows Figure 3 depicts the combined wind event size, wind event duration, and wind quiescent event for "raw" data. Measurements were made at about 40 weather stations with different lengths of the record but for a common recording interval of 180 minutes. All plots clearly show power-law dependence which confirms that wind over the Antarctic is self-organized criticality and also that wind circulation shows ergodic behaviour. "Raw" wind data presented on Fig. 3, were ranked and binned. The following scaling parameters were obtained: $\alpha_s = 1.06 \pm 0.01, \alpha_t = 1.49 \pm 0.04$ and $\alpha_q = 2.26 \pm 0.05$.

## 4. McMudro Dry Valleys Revisited

In December 1903, Capitan Robert F. Scott and his team descended from Victoria Land Plateau toward McMurdo Sound *via* one of the glaciers leading to the sea. While descending Captain Scott observed: "At the end of the second lake, the valley turned towards the north-east … Quite suddenly these moraines ceased, and we stepped out onto a long stretch of undulating sand traversed by numerous small streams, which here and there opened out into small, shallow lakes quite free from ice.

I was so fascinated by all these strange new sights that I strode forward without thought of hunger … We commanded an extensive view both up and down the valley, and yet, except about the rugged mountain summits, there was not a vestige of ice or snow to be seen; and as we ran the comparatively warm sand through our fingers and quenched our thirst at the stream, it seemed almost impossible that we could be within a hundred miles of the terrible conditions we had experienced on the summit … It is certainly a valley of the dead; even the great glacier which once pushed through it has withered away."[16]

However, since the original discovery by Captain Scott, some simple questions remain unanswered. In particular, there is the question of why the adjacent glacier to the south (the Ferrar) reaches the sea? The Ferrar has a less important connection with the inland ice (*via* the South Arm) than the Taylor Glacier. The Ferrar, however, receives several tributary glaciers, notably the Emmanuel, Descent Pass and Overflow Glaciers. The explanation may be that the Taylor Glacier occupies a "locally dry" area. These are rain-shadow "deserts". It is conceivable that a similar Föhn wind operates in Victoria Land, but data are not available."[17]

Much effort has been devoted to understanding the physical, chemical, and biological processes in the McMurdo Dry Valleys (77°30'S, 63°00'E); the largest relatively ice-free area on the Antarctic continent. This ice-free area is approximately 4,850 sq km.

The low precipitation relative to potential evaporation, low surface albedo, and dry katabatic winds descending from the Polar Plateau result in extremely arid conditions at the Dry Valleys. In spite of many papers questioning the atmospheric air movement in the McMurdo Dry Valleys, a recent publication[18] attributed wind events at these valleys as a Föhn (Foehn) instead of katabatic winds.

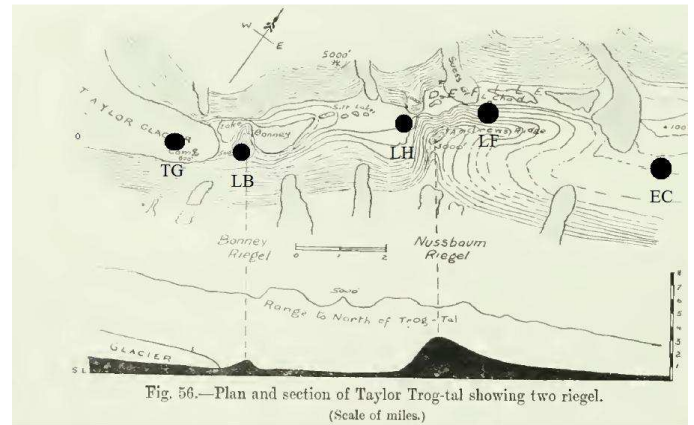

Figure 4. Figure taken from Griffith Taylor's book on the physiography of the McMurdo Sound with a plan and section of Taylor Valley located in the McMurdo Dry Valleys area. Modern automated weather stations are indicated by black points on the figure. The scale is given in miles

It is widely accepted that katabatic wind is the wind that carries high density air from a higher elevation down a slope under the force of gravity. It is also customary to call dry down-slope wind that occurs in the lee (downwind side) of a mountain range, Föhn wind.
Griffith Taylor was one of the more insightful members of Captain Scott's expedition. Figure 4 which is taken from Taylor book[19], illustrates the general settings of one of the Dry Valleys bearing G. Taylor's name. This particular valley is the most studied one. Some time ago several automated weather stations were installed by American scientists[20] at the locations approximately indicated on Taylor's figure.

In the light of the results presented in previous sections of this contribution, it is not surprising that the winds in Taylor Valley are also self-organized criticality phenomenon. Figure 5 depicts cumulative size distribution function $p_s(w_s)$ for "raw" data for each automated weather station.

After performing binning, scaling parameters have been calculated with the results collected in Table 1. The distribution of scaling parameters is interesting.

Table 1. Taylor Valley (McMurdo Dry Valleys) power-law scaling parameters for wind event size $\alpha_s$, duration of wind event $\alpha_t$ and quiescent wind event $\alpha_q$, respectively.

| Location & AWS Station | Elevation [m] | $\alpha_s$ (size) | $\alpha_t$ (duration) | $\alpha_q$ (quiescent) |
|---|---|---|---|---|
| Taylor Glacier | 334 | 0.96 ±0.02 | 1.08 ± 0.03 | 3.46* ± 0.11 |
| Lake Bonney | 64 | 1.24 ± 0.01 | -1.36 ± 0.02 | 3.79* ± 0.10 |
| Lake Hoare | 78 | 1.15 ± 0.01 | 1.23 ±0.02 | 4.55* ± 0.16 |
| Lake Fryxell | 19 | 1.35 ± 0.01 | 1.47 ± 0.03 | 4.83* ± 0.23 |
| Explorers Cove | 26 | 0.98 ± 0.01 | 1.09 ± 0.02 | 4.69* ± 0.08 |

*Asterisk denotes values obtained from small number of quiescent data.

I have already described the general frame of air transport over the Antarctic continent including the Polar Vortex. The following spatiotemporal picture of air circulation in Taylor Valley emerged:
- Transported within the Polar Vortex air is cooled due to radiative cooling,

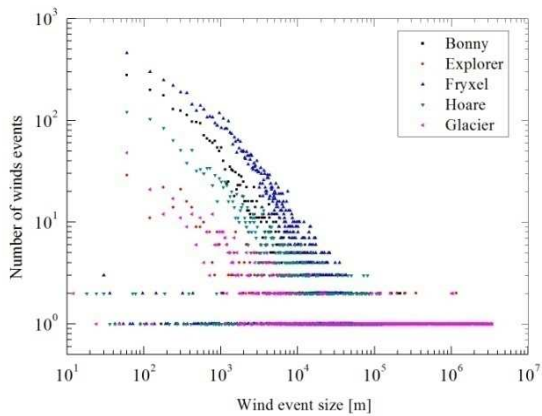

Figure 5. Cumulative size distribution function of wind event size recorded in Taylor Valley at different locations indicated in Fig. 4.

- Since the density of air is inversely proportional to its temperature the air will flow downwards along the Antarctic Plateau toward the Taylor Glacier and the Taylor Glacier weather station,
- Provided that air flow is fast enough, the air flowing downwards would adiabatically warm up,
- Adiabatic transformation takes place when the transformation is fast enough so that the total heat of air is conserved during the transformation,
- The first noteworthy air flow disturbance takes place at the about 500 feet (152 m) high Bonny Riegel (see Fig. 4),
- Depending upon Froude number ($F_r$) flowing air will dam up against Bonny Riegel or will cross it,
- If $F_r < 1$, the katabatic air flow will undergo a (minor) Bonny orographic lift,
- If the contrary, the katabatic flow will overpass Bonny Riegel and continue toward the next obstacle, which is the Nussbaum Riegel at about 3000 feet (914 m) high (see Fig. 4),
- If $F_r < 1$ a (major) Nussbaum orographic lift of katabatic air flow would be observed,
- Orographic lift leads to air masses gaining altitude,
- Due to decreasing atmospheric pressure with increased altitude, the air mass adiabatically cools down,
- The relative humidity may also rise as the air approaches dew point,
- The katabatic air moisture condenses onto the Nussbaum Riegel slope and eventually precipitates on the katabatic-ward side,
- Depending upon Froude number and the mass of damped air, the hydraulic jump (lift) is possible,
- On the leeward slope of Nussbaum Riegel, descending air is dry, and proper Föhn wind is observed,
- Due to different adiabatic lapse rates of moist and dry air, the air at the leeward slope of Nussbaum Riegel becomes warmer,
- From the leeward slope, the Föhn wind moves toward the ever widening Taylor Valley with its exit jet produced by apparent differences of air pressure.

The processes described above are reflected in the changes of the scaling parameters at different locations within Taylor Valley as presented in Table 1. Further data collection from the present weather stations and possibly a few additional ones, would enable the possibility of using the Taylor Valley for modelling meteorological events. A combination of a self-organized criticality with a devil's staircase type geomorphology analysis would generate new and interesting results.

## 5. February 27-March 19, 1912 – Extreme Cold Snap

Captain Robert F. Scott's name is permanently written into the history of Antarctic discovery. His two expeditions: *Discovery Expedition* (1901-1903)[21], and *Terra Nova Expedition* (1909-1913)[22] marked by the ultimate death of Capitan Scott and his four companions left posterity with many unanswered questions.[23]

One of the most perplexing of these questions is the one concerning the events from the final days of the struggle of Captain Scott and his companions to the Cape Evans home base en route from the South Pole.

Life in Antarctica is inevitably related and dictated by the weather, its modes, severity, and apparent unpredictability. Then and now, a small weather window of opportunity appears for humans to explore the vast Antarctic continent. The size of this opportunity is simply measured by air temperature, human endurance and mileage to be crossed. If the distance travelled was to be extended, it meant endurance and organization had to be better, or as one would say better performance on a personal and organizational level would be needed. The game to reach the South Pole was a performance game. We know the results. Amundsen first, Scott second. We know or at least assume to know that Roald Amundsen's trip to the South Pole was a flawless effort, and Captain Scott on the contrary, suffered on many occasions.

However, as far as the Amundsen and Captain Scott expeditions and reaching the South Pole is concerned, it is useful to make a *ceteris paribus* assumption. The assumption is that, in spite of different methods, means, and human effort, both expeditions were able to reach the South Pole and both teams were capable of returning safely to the base camp at Framheim and Cape Evans (Hut Point), respectively. This leaves the weather, understood as a combination of temperature and wind speed, as the only *independent* variable.

I just mention two related examples. The first is Amundsen's premature attempt to reach the South Pole on 8 September 1911, which had to be abandoned due to extremely low temperatures. The extremities of these temperatures are not due to any *absolute* temperature values, but because the dogs could not withstand the impact.[24] The second example is the four day blizzard, which struck Captain Scott's South Pole party at the foot of the Beardmore Glacier on December 5, 1911.

After examining the weather record of the Amundsen and the Scott South Pole parties a list can be presented of temperature and wind speed records measured on every day of their trip. Just by looking at the respective plots of temperature and/or wind speed versus time, rather regular (expected) variations of these variables can certainly be observed. Even the just mentioned four-day-blizzard would fall into this ordinary category. Only two meteorological events described by Captain Scott would fall into the category of *extremely rare* weather events, sometimes called in power-laws jargon[25]: a *black swan*, *outliers*, *dragon-kings* or rogue wave (solitary wave), as observed by Ernest Shackleton just before landing on the south shore of the South Georgia Island.[26]

These black swan events described by Captain Scott, followed each other. The first black swan event occurred on a February 27-March 19, 1912 – Extreme Cold Snap and the second occurred on a March 20-29, 1912 – Never Ending Blizzard.

In this and the following section, I use modern mathematical tools and methods to examine the possibility of these two black swan events happening back in 1912. Essentially, I will look at the possibility of predicting weather variables at different locations at the Ross Ice Shelf using as a reference, the respective variable measured at Ross Island.

Thanks to a fair amount of data accumulated during the period of 1985-2011 by automated weather stations at the Ross Ice Shelf, it is possible to analyze the meteorological record of the Captain Scott expedition.

In my analysis, I use two sets of daily *minimum* near surface air temperature data measured at various geographical locations at Ross Island and at the Ross Ice Shelf (Fig.6). These sets of data were collected during the years 1911-1912 and 1985-2009. In the first case, temperature data were measured by the members of the British Antarctic Expedition 1910-1913 (*Terra Nova*) under the command of Captain Robert F. Scott. In the second case, respective minimum near surface temperature data were measured by modern automated weather stations.

All pertinent aspects of meteorological measurements, data and discussion were collected in a three volume treatise by George Simpson.[27] He was *Terra Nova* expedition's chief meteorologist. Historical temperature data constitute a subset of meteorological records. These records were collected at permanent weather stations and at the routes travelled by Captain Scott. Approximate locations of land weather stations (historical and modern) are clearly depicted on Fig.6. During the *Terra Nova* expedition, the land based meteorological measurements were taken at Cape Evans of Ross Island. The measurements were taken every hour by Simpson and his Canadian assistant, Charles Wright. The temperatures were taken in the screen mounted behind the expedition hut, about five feet above ground on *Windvane Hill*. Four thermometers were placed in the screen: a mercury dry bulb thermometer, a mercury maximum thermometer, a spirit minimum thermometer and a bimetallic thermograph. The measurements were taken in air-free conditions. Before and after the expedition, Simpson ensured adequate testing procedures.

The sledging parties did not take hourly measurements for obvious reasons. Usually three recordings were taken: in the morning, at lunch time and in the evening. Some sledging parties, for example the Main Polar Party (Captain Scott party), were carrying so-called minimum temperature thermometers, in addition to regular thermometers. Captain Scott's party used high quality thermometers, calibrated at Kew Observatory, London. Sling and dry-bulb thermometers were used with precision, and measured at about a $±0.5°F$[28] uncertainty. A specially constructed sling thermometer with a wooden handle was broken by Lt. Bowers on March 10, 1912. From that day on, only Captain Scott's personal spirit thermometer data were available.

Each automated weather station measures wind speed, wind direction, temperature, and atmospheric pressure. The wind speed, wind direction, and temperature gauges (sensors) are mounted at the top of the tower, at a nominal height of 3.9 m. Station atmospheric pressure is measured at a nominal height of 1.5 m. The heights of the gauges (sensors) may change due to snow accumulation at the site. Measurements from the sensors are made every 10 minutes and are transmitted *via* the ARGOS data collection system and processed at the University of Wisconsin. A semi-automated quality control process is applied to 10-minute data. Untreated data are also available. Hourly observations are cre-ated using the closest valid observation within 10 minutes of the hour from the quality control processing.[29]

To perform retrodiction of the daily minimum temperature at the geographical locations of the weather stations, I have selected a back-propagation neural network. It is a network which is able to train the output (Linda, Schwerdtfeger, Elaine minimum temperature data) units to learn to classify patterns of inputs (McMurdo minimum temperature data).[30] There is no theoretical prescription for the number of hidden layers. I have found that a fully connected back-propagation neural network as depicted on Fig. 7 performed very well for the temperature data studied in the present paper. The neural network used in this paper was a modified version of the neural network used by me in previous studies of solutions of the quantum mechanical Schrödinger equation.[31] While experimenting further with different configurations of the network I have found that a network with 3 input neurons and 4 neurons in each first and second hidden layer and 1 output neuron, gave the best retrodiction results. The neural network was trained on sets of daily minimum temperatures in the following way. I presented to the input neurons, a sequences of equally spaced samples of McMurdo daily minimum temperatures $T_{min}(t), T_{min}(t+1), T_{min}(t+2)$, where t stands for a particular Julian day considered in this work period of time. The output neuron was assumed to be the desired retrodiction daily minimum temperature $T_{min}^*(t+3)$ from the respective Linda, Schwerdtfeger and Elaine data. After the network reached its maximum performance for the given set of data, I shifted $t \rightarrow t+1$ and repeated the learning procedure. Thus, in such a way I have obtained three fully trained neural networks for the retrodiction of minimum daily temperatures at geographical coordinates of the mentioned weather stations.

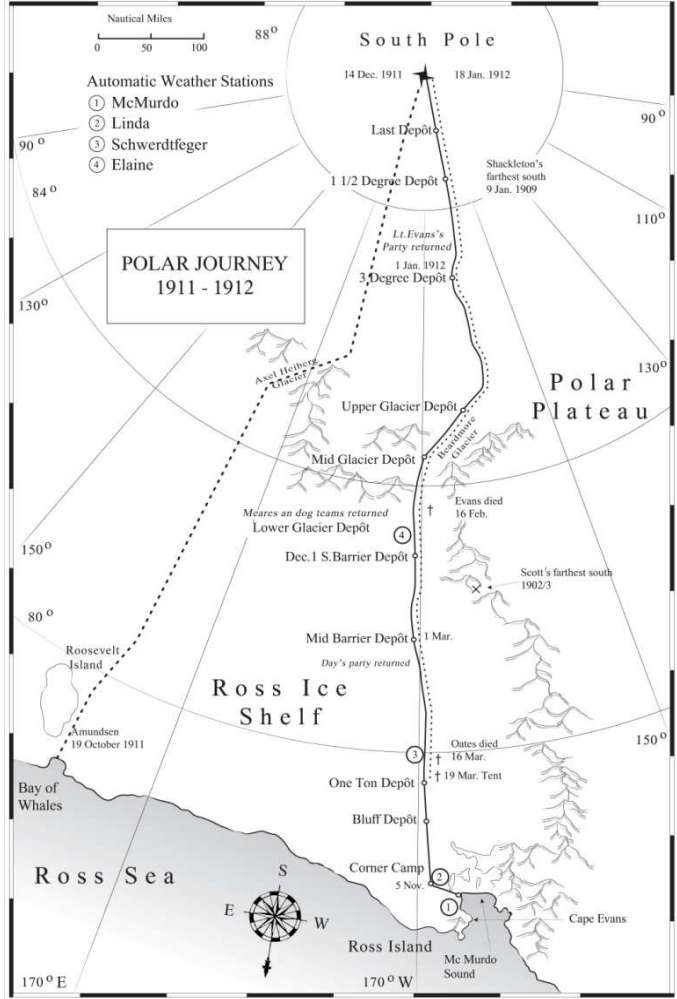

Figure 6. Approximate drawing of the Antarctic route travelled by Captain Scott (solid and dotted lines), and the auxiliary and relief parties in 1911 and 1912, which travelled essentially the same routes as Captain Scott's party. Positions of automatic weather stations are also shown. See the Figure Legend for additional information.

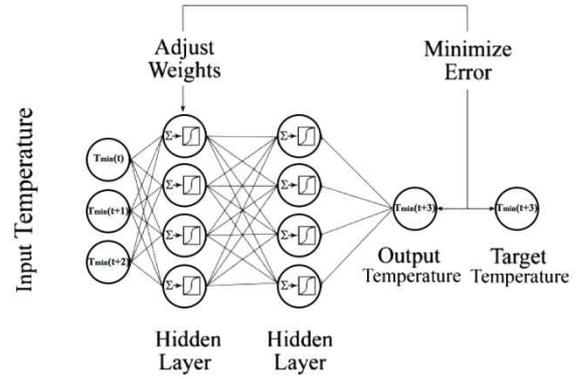

Figure 7. Architecture of back-propagation artificial neural network.

The mean *absolute* retrodiction error $\langle \varepsilon \rangle$ was calculated from

$$\langle \varepsilon \rangle = \pm \frac{1}{N} \sum_{j=1}^{N} \frac{1}{n} \sum_{i=1}^{n} \left| T_{AWS}^{(i,j)} - T_{ANN}^{(i,j)} \right| \qquad (7)$$

where N is the number of years, n is the number of days. The minimum daily temperatures at automated weather station (AWS) and retrodicted by artificial neural network (ANN) are denoted respectively. The error $\langle \varepsilon \rangle$ is not the standard deviation of the sample of retrodicted minimum temperatures. It is the mean *absolute* retrodiction error.

Automated weather stations are localized in the proximity of Captain Scott's route and expedition depots: Corner Camp (Linda), One Ton Depôt (Schwerdtfeger), and South Barrier Depôt (Elaine) as depicted in Fig.6.

After selecting architecture for the best performing neural network I have examined its retrodiction performance by sequential deselecting of yearly minimum temperature data from the training data series. This was done for the McMurdo, Schwerdtfeger stations and performed retrodiction of deselected data was made. Artificial neural network performed extremely well in retrodiction of the minimum temperatures for 25 consecutive years of modern data. The performance of the artificial neural network is described in the following section. I have calculated the mean absolute retrodiction error of the sample of retrodicted minimum temperatures from February 27 to March 19 for each year. The result being $\langle \varepsilon \rangle = \pm 7.1°F$. This indicates a fairly acceptable precision of the retrodiction power of my artificial neural network.

Figure 8 depicts the averaged daily minimum temperature change from February 1 to March 19, at the McMurdo, Schwerdtfeger and Elaine weather stations. Even though these two latter stations are approximately 229 and 588 km apart from McMurdo, one can easily notice that some unspecified relationship of minimum temperature changes between these stations is present. A gradient in the minimum temperature at one station is followed by a similar change at another station and/or *vice versa*. Thus, the components of the gradient of minimum daily temperature, transform covariantly under changes of coordinates i.e., the geographical coordinates at the Ross Ice Shelf. I do not imply that there is a mathematical rigorousness of the mentioned covariant transformation and/or linear relationship.

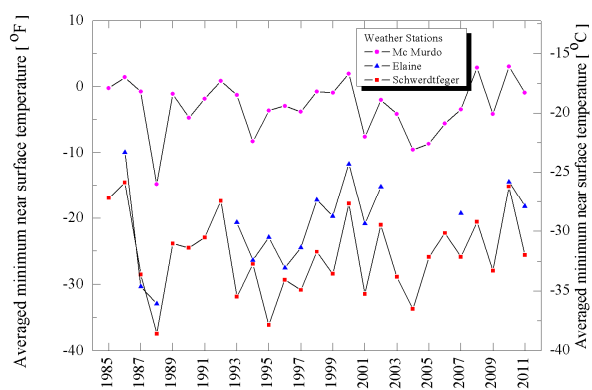

Figure 8. The minimum daily near surface temperature averaged (the arithmetic mean) from the 1985-20011 McMurdo, Schwerdtfeger and 1993-2011 Elaine weather stations.

Although averaged minimum temperatures presented in Fig. 8 can be exceptionally well approximated by linear regression analysis, its very nature is nonlinear. One has to notice that I have presented only temperatures for the first quarter of the year. The actual yearly averaged minimum temperature has a U shaped letter with a distinctive coreless winter, April through September and a short-lived crest temperature between the beginning and end of summer, December through January. Fig. 8 also shows that although the Elaine station is further South than the Schwerdtfeger station, the daily minimum temperature at the Elaine is frequently close to or above that of the Schwerdtfeger station. I attribute this phenomenon to an adiabatic effect of air warming by katabatic winds flowing downwards from the Beardmore Glacier. With this in mind one can further confirm a mirrored similarity of the daily minimum temperatures changes between McMurdo, Schwerdtfeger, Elaine and Linda (which is not shown for clarity) weather stations. I attribute this "*mirrored similarity*" to the essentially flat surface of the Shelf and to the prevailing south and southeast by south winds that are directed along the Transantarctic Mountain pathway. Therefore, a mirrored similarity of minimum temperatures along the route of Captain Scott and the auxiliary parties is self evident.

I could have advance my study by using the above mirrored similarity. It seemed essential in the analysis, however, to account for fluctuations and nonlinear trends of minimum temperature changes as a continuous-state, discrete-time stochastic process. For this purpose I have selected an artificial neural network for a time series prediction and retrodiction of minimum temperature data. After selecting architecture for the best performing neural network I have examined its retrodiction performance by sequential deselecting yearly minimum temperature data from training data series. This was done for the McMurdo - Schwerdtfeger stations and performed retrodiction of deselected data was made. From Simpson's experimental studies and report we know that the uncertainty of thermometers used by Captain Scott's expedition was about ±0.5°F. It is roughly the size of points indicating temperatures changes in the figures in this paper. From a possible 25 year set of training temperature data for the McMurdo-Schwerdtfeger stations I have selected a 24 year set of training data and performed retrodiction for the not selected for training 25th year. Thus for each deselected year I calculated the respective *absolute* retrodiction error for retrodicted minimum near surface temperatures, for the days from February 27 until March 19. In this way I have obtained *absolute* retrodiction error for each deselected year; all together 25 *absolute* retrodiction errors. The mean *absolute* retrodiction error is $\langle \varepsilon \rangle = \pm 7.1°F$.

The measured and reported minimum temperatures by Captain Scott's party together with retrodicted values in the vicinity of the Schwerdtfeger weather station (One Ton Depôt, the final miles and the last of Captain Scott's camps) form the historical data of minimum temperatures measured at

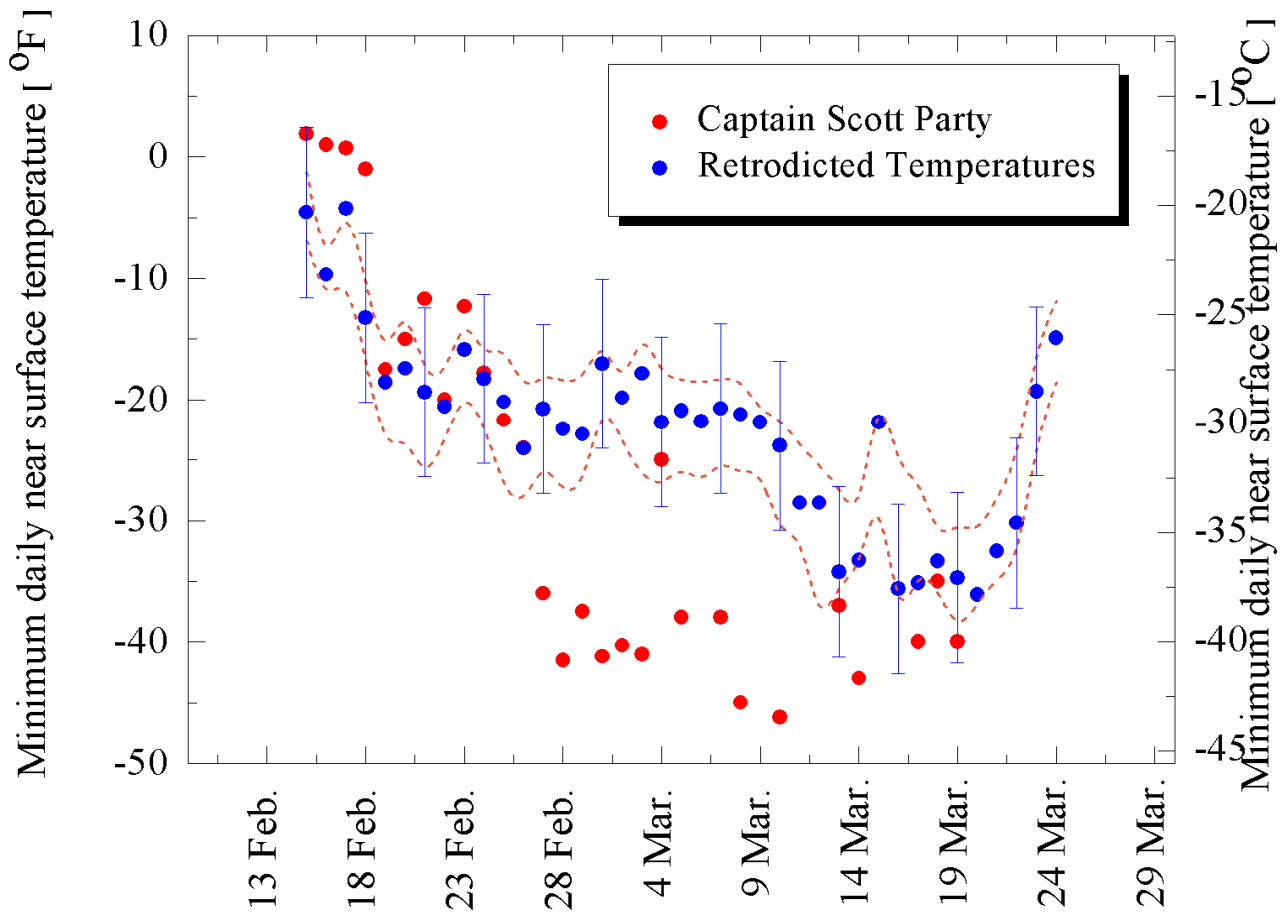

Figure 9. Historical near surface minimum daily temperatures (or the *lowest*) reported by Lt. Bowers and Captain Scott, together with respective minimum temperatures retrodicted from historical minimum temperatures recorded at Cape Evans, in 1912. Temperature readings after March 10 are not minimum near surface temperatures.

Cape Evans. They are clearly depicted in Fig.9. It is not difficult to note that until February 27, 1912 the retrodicted and the temperatures reported by Captain Scott confirm anticipated short lived random fluctuations. Compared to the temperature data presented in Fig. 8 for the Schwerdtfeger and Elaine stations, Captain Scott's reported minimum temperatures were slightly above retrodicted temperatures for the geographical position of One Ton Depôt. This is to be expected as his party was South from that depôt between February 15-23. It should also be noted that until the latter part of February the formerly observed mirrored similarity, temperature reflection was well established.

However, from February 27 on, for twenty consecutive days (with an exception of March 4), reported and retrodicted minimum daily temperatures rapidly and significantly diverge. Retrodicted minimum near surface temperature, is on the average -13°F above that reported by Captain Scott and his party.

The discrepancy is significant in value and length. Such a discrepancy can be attributed to:
(i) inaccuracy of my retrodiction method,
(ii) incorrect temperature data readings due to Captain Scott's party thermometers malfunction,
(iii) long term temperature change in the Antarctic and/or
(iv) the fact that temperature data documented by Lt. Bowers and Captain Scott were distorted to exaggerate real weather conditions.

In the above analysis and discussion I have clearly confirmed that I was able to retrodict historical minimum temperature data reported by various parties. Narrowing down to the weather conditions at One Ton Depôt, it can be argued that because my data for the training of a neural network at the Schwerdtfeger station did not contain freak temperature events, I could not retrodict similar events in the past. However, this is not the case. The training set of temperature data contained all possible temperature recordings and the observed minimum temperatures fluctuated in the wide range of $\langle +17.8 \div -68.8 \rangle$°F. Therefore, I would argue that all minimum temperatures, except Captain Scott's temperature data, measured at the environs of the final camp, show that the Main Polar Party reported temperature data that are in dispute with all the remaining temperature data. In the case of the Schwerdtfeger station, for retrodiction I have used twenty seven years (1985-2011) of solid temperature data which contain a great number of minimum temperature fluctuations. More importantly, only temperature data from Captain Scott and his party for late February and March diverge from the established trends and do not reflect the *mirrored similarity* of minimum temperatures measured at Cape Evans by Simpson and Wright. It is also pertinent to note that it is a remarkably long time to have twenty consecutive days of divergence in minimum temperatures. This alone excludes the commonly held notion of a rare event occurring in the system which constantly fluctuates around its mean. In my case it is the arithmetic mean of minimum daily near-surface air temperature. Estimating small probabilities or rare events is a considerable challenge for heavy-tailed distributions which do not have exponential moments. However, all the automated weather stations under consideration, smoothly passed the Shapiro-Wilk test of the null hypothesis for minimum temperature data. This confirmed normal distribution of temperatures with exponential moments. Additionally, the possibility of a consecutive twenty days of rare minimum temperature events is even more unlikely.

I have already mentioned that in training and followed retrodiction calculations I have used temperature data collected at the McMurdo weather station. I have pointed out that this station is about 26 km from Captain Scott's Hut at Cape Evans where actual historical measurements were taken. Although there is no satisfactory historical or modern evidence one could imply that there could be a possibility of temperature difference between these two sites. My neural network is ideally suited to answer any question as to how eventual temperature difference would affect overall prediction and retrodiction procedures related to discussed case. I have performed a numerical experiment by training a neural network with minimum temperature data different from original McMurdo data by ±3.6ºF (±2ºC). After that I have used historical data to retrodict minimum temperatures at One Ton Depôt. The results are depicted on Fig. 9. The upper dashed curve was obtained after the McMurdo data were perturbed by +3.6ºF and the lower dashed curve by -3.6ºF, respectively. This numerical experiment shows that even if there was slight temperature dependence due to intricate meteorological and/or physical features at Ross Island it was negligible and insignificant in formulating the conclusions of the present work. Simultaneously this numerical experiment further confirms that the artificial neural network used in this work is sensitive and capable of response to fine temperature changes in training and analysis data.

I have already mentioned that the uncertainty of thermometers used by Captain Scott's expedition was ±0.5ºF. The only thermometer left after March 10, 1912 was Captain Scott personal spirit thermometer which was found by a search party in 1913. Charles Wright tested its calibration back in London. Test results proved this thermometer's accuracy within a tenth of a degree.[32]

Although, scientific debate over the issue of global warming is still open, recent results further confirm a positive warming trend in the Antarctic.[33] The warming trend varies with the geographical position over the Antarctic, however for approximate coordinates of the Schwerdtfeger weather station the estimated warming trend is about 0.18ºF (0.1ºC)/decade. I believe that this warming trend does not contribute significantly to my overall analysis and conclusions.

By eliminating the alternative explanations (i), (ii) and (iii) mentioned above, only one conclusion is left (iv). That is, that the temperature data reported by Lt. Bowers and Captain Scott himself in late February and March 1912 were distorted by them to exaggerate and dramatize the weather conditions. My results clearly show that Captain Scott, Dr. Wilson and Lieut. Bower's deaths were a matter of choice rather than chance. The choice was made long before the actual end of food, fuel and long before the end of their physical strength to reach imaginary salvation at One Ton Depôt.

# 6. March 20-29, 1912 – Never Ending Blizzard

The second black swan in the meteorological record of Captain Scott, is a blizzard of unheard proportions. The blizzard started on March 20 and ended on March 29, 1912.

The Captain Scott and Amundsen expeditions were definite events which during the planning and executing stages were reduced into numbers: miles per day, calories per day, temperature, efficiency, work, forecast, friction, etc. Both explorers and the members of their expeditions, in different degrees of proficiency and expertise, transformed these numbers into everyday life and action. Amundsen and Captain Scott had knowledge and practice in polar regions while planning the South Pole expedition. Amundsen gained his skills during the *Belgica* (1897–99)[34] expedition and Northwest Passage (1903–1906).[35] Captain Scott acquired his familiarity with the Antarctic throughout his *Discovery* Expedition (1901-1904).[36]

In modern terms, as we are accustom to think the entire Universe is a digital machine[37], the Captain Scott and Amundsen expeditions may be digitalized and analyzed along these lines. Such an observation may be quite palpable for anyone who has ever planned a vacation trip or even a more challenging self-sustained expedition. Indeed, the Antarctic Manual[38] issued by the Royal Geographical Society was intended as a digital description of various aspects of the planned Captain Scott Discovery Expedition. Unfortunately, instead of promoting "correct digits" the Antarctic Manual was a fairly random collection of different papers on various subjects related to the expedition. Markham was straightforward about his input[39]

> "I planned an Antarctic Manual on the lines of the Arctic Manuals prepared for the expedition of 1875-76, securing the services of Mr. G. Murray as editor. It proved very useful, the first part containing instructions and information by leading men of science, and the second part being the narratives of Biscoe, Balleny, Dumont d'Urville, and Wilkes, with papers on polar travelling by Sir Leopold M'Clintock and on the exploration of Antarctic lands by Arctowski."

More importantly, the Manual was used by Sir Cements Markham to promote his interests by selecting certain contributors rather than a sound scientific and digital handbook.

In particular, an article by Sir Leopold McClintock reprinted in the manual, presented *incorrect values* of sledging dog food rations. The rations advised by McClintock were two times too low. This error of negligence by McClintock, Markham, Murray, Scott and Shackleton pushed the whole history of British Antarctic exploration into the *Heroic* Age of Antarctic Exploration instead of into the *Grand* Age of Antarctic Exploration.

It seems that Roald Amundsen and his team were sufficiently skilled at keeping and following the right numbers. Amundsen, unlike Captain Scott, was able to capture the right digits and the relationships between them. He followed what I call, 'fine tuned polar exploration' which includes subtle digital values and their relationships between nature (physical and life) and human needs and actions.[40]

Although the Amundsen and Captain Scott expeditions were preliminary logistic undertakings, this fundamental aspect of these expeditions was neglected by historians, bio- and hagiographers. Only a few exceptions exist scattered in professional journals. The majority of books concentrated on nostalgic lamenting over the suffering of Captain Scott and his companions. Authors close their eyes to scientific analysis while arguing that Captain Scott's expedition was a preliminary scientific expedition. The problem goes as far as data dragging, and/or attributing exaggerated data (increasing daily temperatures, wind speeds and its duration) to the Captain Scott expedition. In the case of Susan Solomon's work, there even occurred open falsification of data, to prove the author point.[41] However, falsifications or so-called 'minor adjustments', are not a risky business since Leonard Huxley got away with his falsification of Captain Scott's temperature records almost one hundred years ago.

Polar historians and enthusiasts are aware that toward the end of March 1912, Captain Scott, this time without the support of Lt. Bowers or Dr. Wilson, reported in his journal a meteorological event, which was extraordinarily as far as its length and strength was concerned. This event was the gale which lasted nine or ten days. Captain Scott's entry tells us "Thursday, March 29. – Since the 21st we have had a continuous gale from W.S.W. and S.W." And continues "Every day we have been ready to start for our depot *11 miles* away, but outside the door of the tent it remains a scene of whirling drift."[42]

A careful reader would note from Captain Scott's own descriptions or from Simpson's meteorological record and analysis, that frequently observed gales (or blizzards) at the Ross Ice Shelf last about one-two days. One or two days is to be expected, but nine or ten days?

Were the laws of physics suspended at the end of March 1912 in the Antarctic? Did it happen for a second time? Did it happen only at the actual location of Captain Scott's team and did it not occur at locations where simultaneous measurements were taken by other members of the *Terra Nova Expedition*?

I have already shown that the first black swan in the meteorological record (minimum temperatures) of Captain Scott did not happen, and the laws of physics were not suspended on the Ross Ice Shelf in 1912.

Captain Scott addressed a similar question. At the entrance to the Beardmore Glacier, Captain Scott's South Pole party was stopped and held in their tents for four long days by a blizzard or gale as he called it in the *Message to the Public*. Captain Scott posited the question, "What on earth does such weather mean at this time of year?" and wonders: "Is there some widespread atmospheric disturbance which will be felt everywhere in this region as a bad season, or are we merely the victims of exceptional local conditions? If the latter, there is food for thought in picturing our small party struggling against adversity in one place whilst others go smilingly forward in the sunshine."[43]

Captain Scott's reasoning is interesting indeed. It may sound as a beacon to his lines in his famed Message to the Public "our wreck is certainly due to this sudden advent of severe weather, which does not seem to have any satisfactory cause."[44] But in the above, Capitan Scott's comment on two variables: distance and time are also missing. The distance between the entrance to the Beardmore Glacier and McMurdo Sound is about 600 km. It is more than the distance between London and Edinburgh. Does anyone expects or anticipates analogous weather conditions between these two cities?

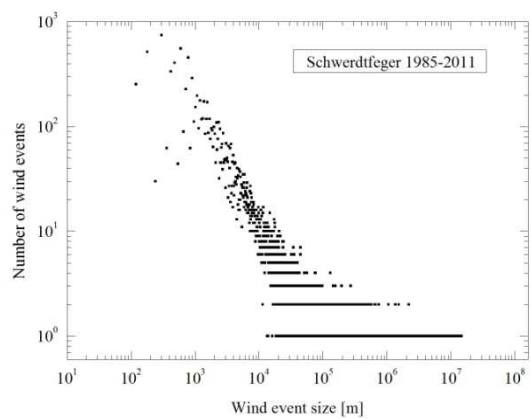

Figure 10. Cumulative wind size distribution function at Schwerdtfeger weather station located near One Ton Depôt and the final camp of Captain Scott's party.

Cutting that distance would be about the position of the One Ton Depôt or Schwerdtfeger (-79.904°,169.97°) automated weather station or Captain Scott's last camp. The very place where the second black swan blizzard struck and held the party in the tent at gale strength, for nine/ten days. Is it really possible that such a gigantic (duration and strength) wind event can actually take place? Is it possible that such a gigantic wind event can take place as geographically isolated phenomenon?

In my analysis, I used 26 years of wind data measured at Schwerdtfeger and at the Antarctica New Zealand Scott Base[45] station located at the southern tip of Hut Point Peninsula of Ross Island, Antarctic. New Zealand's Scott Base is directly "facing" all meteorological events at the Ross Ice Shelf.

Both Scott Base and Schwerdtfeger stations are not exceptions to the self-organized criticality of wind events shown in the above preceding sections. In Figure 10, I have depicted the cumulative size distribution function $p_s(w_s)$ for "raw" data for Schwerdtfeger. Wind event size distribution at this station shows a typical power-law behaviour with scaling parameter $α_s = 1.27 ± 0.01$. Similar wind size scaling parameters calculated for Scott Base and McMurdo stations are $α_s = 1.13 ± 0.01$ and $α_s = 1.31 ± 0.07$, respectively.

Up to now, the belief that arbitrary size wind event may happen in the Antarctic was taken for granted by polar enthusiasts and authors. Although the nine/ten day gale at the end of March 1912, described by Captain Scott, was indeed an extraordinary event, no one bothered to look at meteorological data.

I have shown above, that due to the evident scaling property of wind event size at Schwerdtfeger and the remaining stations/locations, it can be argued that at this particular location, which is very close to the last camp of Capitan Scott's party, *arbitrary* wind event size is certainly possible. Thus, one can attempt make scholarly conjecture based on the above analysis of the scaling properties of wind events, that indeed the nine/ten days gale reported by Captain Scott took place.

However, such a conjuncture is not correct. The occurrence of arbitrary wind event size is only a theoretical possibility that power-law relationship $p_s(w_s) \propto w_s^{-0.75}$ holds at Schwerdtfeger station, without physical limits or underlying physics laws.

On a grander scale than the Schwerdtfeger location, I have shown above, that the wind is self-organized criticality over the whole Antarctic continent. The movement of the wind takes place within the polar cell boundary and the pool (mass) of air transported is limited. Therefore, not arbitrary wind event size can be observed.

Another important and limiting factor is the time needed to pass in order for the observer to note arbitrary small or big event. And the third factor is that wind events at the Ross Ice Shelf are correlated. This means that whatever size wind event occurs at, for example, the Schwerdtfeger location, a similar in size (thought delayed in time) event occurs at the Scott Base and McMurdo station. Alike correlations were also found between near surface air temperatures at these locations.

Present knowledge makes it difficult to determine how overall air pool and time duration of observation may contribute to wind event size at the Ross Ice Shelf. For this reason, I argue that the nine/ten day gale on March 20-29, 1912 reported by Captain Scott, did not take place.

Capitan Scott reported that on March 20-29, 1912, a blizzard of gale force confined the party in the tent. It is fair to assume that Captain Scott, as a Royal Navy officer, was well trained and acquainted in describing wind force according to Beaufort scale.[46] In this empirical measure the wind speed in the range of 17.2–20.7 m/s (62–74 km/h) is named a *gale* – Beaufort number 8. Captain Scott used the term gale or moderate gale 59 times in his journal. Whenever, it was possible I checked Captain Scott description against Terra Nova or Simpson record and I confirmed that Captain Scott description was accurate, with only one exception: the gale of March 20-29, 1912.

The Beaufort scale was a standard for ship log entries on Royal Navy vessels in the late 1830s. However, the Beaufort scale is rather a phenomenological scale.[47] The biggest pitfall of this scale is the lack of a time frame at which a given wind speed has to be sustained in order to fall into a given category. Wind gusts and continuous fluctuations additionally complicate the issue.

Nevertheless, I have assumed that if within a 24h time frame, at least a onetime recorded wind speed was equal or higher than the lower bound gale speed (17.2 m/s=62km/h), then a 24h frame was called "gale-day". If two gale-days were recorded in two, three, … *consecutive* days then such wind event is called a two day gale, three day gale, *etc*.

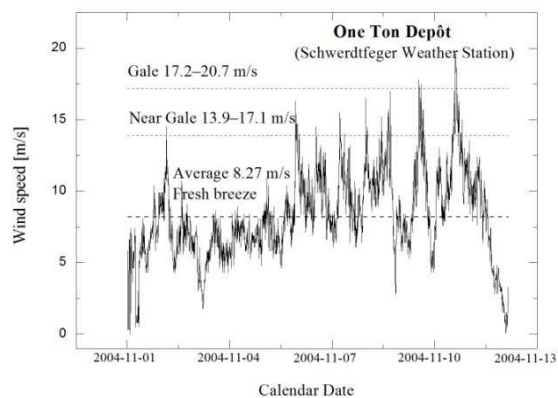

Figure 11. The longest and strongest wind event occurred at Schwerdtfeger station (One Ton Depôt) between Nov. 1 and Nov. 13, 2004.

The general behaviour and structure of wind events have already been presented in Fig.1 for New Zealand's Scott Base. In spite of the erratic behaviour of wind events, a careful examination lead me to conclude that self-organized criticality of wind events is observed across the Antarctic continent. Due to the above mentioned scaling properties of wind events, some assumptions can be made about the longest possible wind event at the location of Captain Scott's last camp in the proximity of One Ton Depôt, and therefore events close to the modern Schwerdtfeger weather station. By taking into account the *whole* available record of wind data 1986-2011. I found that the longest and strongest wind event occurred between Nov. 1 and Nov. 13, 2004. The wind speed structure of this event is depicted on Fig.11. The duration of a wind event is understood as a continuous record of wind speeds $v > 0$, $m/s$ with a detection threshold of about 0.1 m/s.

The Nov. 1 and Nov. 13, 2004 wind event which lasted thirteen consecutive days, had a few important characteristics. First of all, there was extremely fast wind escalation (rise).[48] For an observer from the ground, it appeared as a wind blast. The second important characteristic was the wave structure of the wind speed changes. And the third was the sudden downfall of wind.

To illustrate Captain Scott's wind record, on the same figure I have also shown the lower bound of the Beaufort scale for wind of gale, near gale, and fresh breeze strength. The observation is rather palpable. Captain Scott claimed that at their final camp - about 11 miles from One Ton Depôt, his party encountered a nine/ten day long gale. It is self-evident from Fig.11, that it is far reaching to call this most severe wind event, a gale. There are only two wind *gusts* (spikes) which hardly and briefly reach gale force. Lowering the Beaufort scale to near gale force on the same figure shows that the biggest wind event can hardly be called a near gale event.

At this point, it may be useful to re-call how the Beaufort scale is related to land conditions via the following definitions:[49]

*Fresh Breeze* - Branches of a moderate size move. Small trees in leaf begin to sway,

*Near Gale*- Whole trees in motion. Effort needed to walk against the wind,

*Gale* - Some twigs broken from trees. Progress on foot is seriously impeded.

In the Antarctic, the wind creates near surface drag force. This force drives along air snow particles and sweeps snowflakes which are deposited on the surface into the air. Conditions called a blizzard are then created. Blizzard conditions are those in which visibility and contrast are severely reduced.

According to Captain Scott's last entry in his field Journal on March 29, 1912, the combined conditions of a nine/ten day gale with blizzard conditions were to blame for failure in priority of reaching the South Pole. In particular, Captain Scott's comments "Every day we have been ready to start for our depôt 11 miles away, but outside the door of the tent it remains a scene of whirling drift."[50]

Returning to the biggest wind event presented in Fig. 11, it was rather a fresh breeze which occurred. Such a breeze is much below that of a gale. Thus, the biggest recorded wind event by a modern weather station, at the proximity of One Ton Depôt, was about two times smaller than the event reported by Captain Scott back in March, 1912. It is also important to note that this particular modern record wind event took place in November and not in March.

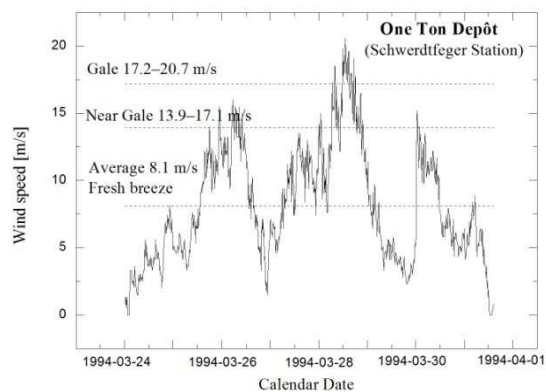

Figure 12. The biggest wind event at Schwerdtfeger weather station (One Ton Depôt), in the month of March. The event was recorded between March 24-31, 1994.

In modern data, it is not difficult to find the biggest wind event at Schwerdtfeger weather station, in the month of March. Such event was observed between March 24-31, 1994 and is depicted on Fig. 12. The presented wind observation in March further confirms the nine/ten day gale described by Captain Scott was much greater in length and force than any wind observed in modern record. This particular wind event also reveals an already described feature of all wind events in the Antarctic. This characteristic is a sudden rise and downfall of wind separated by frequent wind humps. These wind humps are long time scale fluctuations of wind speed.

The definition of blizzard provided by the National Oceanic and Atmospheric Administration[51]

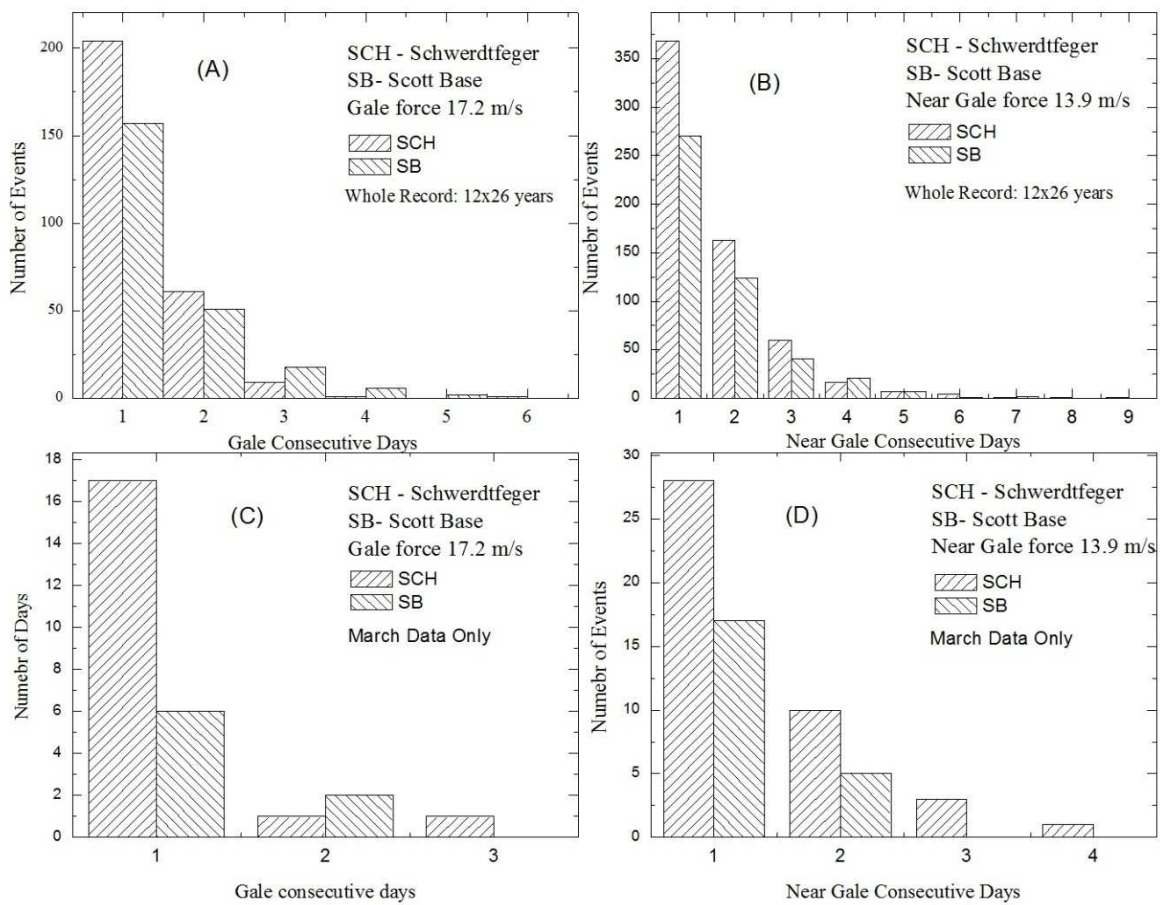

Figure 13. The total number of consecutive gale-days recorded at Schwerdtfeger (One Ton Depôt) and Scott Base weather stations, for whole record and for every month of March from the 1985-2011 period.

requires *sustained* gusts of gale force or greater and blowing (falling) snow which reduces visibility to less than ¼ mile.

Returning to Fig. 11 and 12, it can be noted that within these extreme wind events one can select two different time windows: above the fresh breeze line and below the fresh breeze line. According to this definition, the conditions and the time "spend" by wind events below the fresh breeze line cannot be regarded as blizzard conditions. More importantly, when the weather conditions below the fresh breeze line are favourable, it is thought possible although challenging, to travel. Looking at Fig. 12 it may be concluded that although this longest and strongest wind event lasted about 4 days (96h), there were encouraging weather conditions about half of the time. Thus, the distance of about 22miles could have been crossed at average speed by Captain Scott's sledging party. Obviously, the day & night routine would have to have been changed to a "pitch a tent and go when you can" schedule. Exactly the same observation can also be made by analyzing the biggest recorded November wind event at Schwerdtfeger station (One Ton Depôt). Therefore, Captain Scott's im-mobility due to the nine/ten day gale from March 21-29, 1912 is dubious.

Let us investigate the next question of how many days a gale may last at the Ross Ice Shelf. If within a 24h time frame, at least one time-recorded wind speed was equal or higher than the lower bound of speed of a gale (17.2 m/s=62km/h), then I called the 24h frame a "gale-day". If two gale-days are recorded in two, three, … *consecutive* days, then such a wind event is called a two day gale, three day gale, *etc*. A similar assumption is made in relation to weaker, near gale (13.9 m/s=50 km/h) wind events.

Figure 13 depicts the total number of consecutive gale-days recorded at Schwerdtfeger and Scott Base weather stations, for every month of March from the whole 1985-2011 period. A wealth of knowledge is depicted in this figure. From my our point of view and in relation to Captain Scott's record, it can be seen that in all the analyzed months of March (Figure 13, C & D), the longest consecutive gale and near gale winds lasted for three (3) and four (4) days. That is *far* below the black swan, nine/ten day gale reported by Captain Scott. Fig-ure 13 provides the second argument against Captain Scott's record of a nine/ten day long gale in late March 1912. The gigantic gale winds (and thus, the blizzard) reported by Captain Scott is of unfounded proportions and the laws of physics were not suspended.

I presented the above two powerful arguments against Captain Scott's claim of an endless gale at the end of March 1912. Actually, the gale was not the first claim of unexpected weather by Captain Scott. It is the second unfounded reason used by Captain Scott to explain "the storm which has fallen on us within 11 miles of the depôt at which we hoped to secure our final supplies."[52]

But the One Ton Depôt was not exactly salvation. The Depôt was not adequately stored, because the written orders of Captain Scott had not been followed. The salient mutiny was already taking place at Cape Evans in early 1912. Everyone had their own excuses for leaving the place. The dog driver, Cecil Meares, left on his own terms, G. Simpson the meteorologist, was apparently re-called for more(?) important duets in India, and Atkinson unwillingly assumed the command post without authority nor the means to enforce it. But it was Dr. Atkinson, who through his meteorological record of the rescue attempt in late March 1912, provides the final argument that Captain Scott forged his meteorological record.

Not long after Cherry-Garrard and Demetri's premature dispatch of a relief attempt[53], a decision taken to dispatch the Second Relief Party. On March 26, 1912, Atkinson and Keohane started out alone from Hut Point to search for and help Scott's party. Neither of them could handle dog sledging,
Table 2. The weather Register of the Second Relief Party – Atkinson and Keohane.

| Date Time | Dry Bulb Temperature [°F] | Wind Force | Minimum Temperature [°F] |
|---|---|---|---|
| March 27 14:30 | +2.5 | 3 | |
| 17:30 | -3.5 | 3 | |
| March 28 7:00 | -6.5 | 2 | -6.0 |
| 12:30 | -15.5 | 3 | |
| 17:00 | -6.5 | 1 | |
| March 29 7:00 | -3.5 | 1-2 | -13.0 |
| 12:30 | -0.5 | - | |
| 17:30 | -0.5 | 1-2 | |
| March 30 7:00 | -8.5 | 1-2 | -16.0 |
| 12:00 | -3.5 | 3-4 | |
| 17:30 | -5.5 | 3-4 | |
| March 31 7:00 | -13.5 | 3-4 | -13.0 |

and they resorted to the standard man-hauling technique. Their progress was very slow, only nine miles per day. The party travelled only to a point eight miles south of Corner Camp where Atkinson recorded "At this date [March 30, KS] in my own mind, I was morally certain that the party had perished,"[54] Atkinson conjuncture is surprising, both because of its appeal to moral issues and its *post factum* character. An elementary calculation of the possible arrival time of Captain Scott's party, with a number of updates resulting from information provided by returning parties would give to Atkinson a date March 30 (Corner Camp)[55] which clearly questions his moral confidence.

After leaving Hut Point, Atkinson and Keohane's party was fully exposed to air stream flowing northward along the Transantarctic Mountains.[56] Due to the already shown high correlations of temperatures and wind speeds along Captain Scott's route, including Atkinson and Keohane's course, both parties should have experienced the same or very similar weather conditions.

In the small Table 2 I have presented the weather Register of the Second Relief Party. It is self-evident from this table that the weather conditions recorded by Atkinson were mild. On March 26, Atkinson observed "The temperature was exceedingly low but the weather fair."[57] From March 27 until March 30, he recorded a wind force between 1 to 3. Thus, the strongest wind observed by the relief party was a gentle breeze; leaves and small twigs constantly moving according to the Beaufort scale description.

Comparing these conditions and the analysis of wind duration and strength as a raging gale, at One Ton Depôt, the conclusion is that Captain Scott's wind record was highly inaccurate. The second black swan meteorological event did not take place.

In the above two sections, I presented a detailed analysis of two black swan meteorological events reported by Captain Scott. In both cases I show that events described by Capitan Scott did not take place. The conclusion is a confirmation of my previous inferences. Therefore, I conclude that the deaths of Scott, Wilson and Bowers were a matter of choice rather than chance. It was a choice made long before the actual end of food, fuel and physical strength to reach imaginary and delusive salvation at One Ton Depôt. Captain Scott's reported February 27-March 19, 1912 – Extreme Cold Snap and March 20-29, 1912 – Never Ending Blizzard were invented meteorological events and never occurred.

## 7. Hornsund in Spitsbergen Island Revisited

The Polish Polar Station established in 1957 is located on the fringes of the slopes of Hornsund, on the western side of the southernmost tip of Spitsbergen island. Spitsbergen is the largest island of the Svalbard archipelago.[58] A large human and financial effort was devoted to the study of many different scientific issues of the station area. In particular meteorological measurements and observations played and will play important roles in the climate analysis of the region. In this section, I will briefly show that self-organized criticality wind regime is also observed in the Northern hemisphere. As an example, I have used wind data from the Polish Polar Station. In the following section, I have formalized and generalized the weather analysis methods of Fedorov and Ferdynus. I have shown the methods are a sub-class of rank-ordered occurrence frequency distributions explained by power-law distributions.

In the previous sections, I presented and analyzed the wind events of the Antarctic continent, and I have shown that wind event size, wind event duration and wind quiescent event are all well described by power-law. With this information in mind, one would observe that a fractional Fokker-Planck equation for the probability distribution of air particles whose motion is governed by a non-Gaussian Lévy-stable noise, would be a suitable theoretical approach in the description of near surface winds over the Antarctic continent.

The case of self-organized criticality of near surface Antarctic, katabatic winds seems to be well proven, as shown above. In the case of fundamental universality of physics laws, it may be assumed that similar wind events in similar conditions should be the same. Thus, a katabic winds in the Northern Hemisphere (Greenland, Spitsbergen, … ) and at ice caped planets like Mars should behave in the same way. This means those winds behave as described by self-organized criticality meteorological phenomenon presenting a critical self-organized behaviour, a property of dynamical systems having critical points as attractors.

While thinking about meteorological variables, we are accustom to assuming that variables like temperature, pressure, *etc*. are *bounded* continuous random variables. These variables are indeed continuous. Due to measurement periods and instrument thresholds, we quantize these variables in the form of a catalogue of events or simply a record. In meteorology, we are essentially working with *occurrence frequency distributions*. These distributions are of great importance in meteorology.

In particular, these occurrence frequency distributions are time and value dependent. The time dependence is simply related to Earth-Sun gravitational interactions and solar radiation. However, even at this level, the time dependence of meteorological variables may be complex and fully understood. Climatic classification may also be based on water[59] and the energy-mass budget.[60]

But the most comprehensive method of studying the most common combinations of meteorological variables was proposed and developed by a Russian named E. E. Federov. In his paper entitled Das Lima Ales Wettergesamtheit (Climate as Totality of the Weather), Federov explained:[61]
"The author [Federov] shows that the usual method of presenting climatic data by means, averages, extremes, and sums of the various meteorological elements separately is untrue to nature; for these elements occur and act in combination only."

Federov introduced the "weather case" - the weather that occurs at a given location within a 24h time span. "The climate then consists of a succession of weather cases, which are to be classified into weather types."[62] In order to describe weather cases, Federov divided weather variables into 12 separate states describing temperature, cloudiness, wind and precipitation. Federov's complex climatology was criticized by Nichols[63] who observed that in order to account for true climate description using complex climatology, one has to deal with plethora, $110\times10^6$ possible combinations of letters assigned to weather cases.

Thus, Federov's idea suffered from the significant drawback of a lack of computational power available in 1927. The curious reader may notice at this point, that similar great ideas of Bjerknes and Richardson endured similar computational shortages. This meteorologist attempted to overcome their insufficient computational power by introducing what one would call, compartmental modelling.

In spite of some unsubstantiated criticism that complex climatology "is not a research method"[64], Federov's idea was used in several studies. In particular, Ferdynus, using previously suggested complex climatology, presented the most comprehensive analysis of Hornsund climatology. His picture depictions of dominant weather types at Hornsund are of equal beauty to the famed Mandelbrot fractals or Lorentz weather attractor.

Before proceeding further, I would re-cast complex climatology analysis into modern mathematical terms of marginal distributions. which would serve us in calculations of occurrence frequency distributions of weather. Having done that, I will use marginal distributions to show that one can use them as an occurrence of marginal distributions to show that the weather parents are indeed self-organized criticality phenomenon. The established link between marginal distributions and self-organized criticality and thus the power-law, will show how this phenomenon by simple property of invariance under aggregation arises in meteorology.

Given the distribution of a vector of random variables, it is possible in principle to find the distribution of any individual component of the vector, or any subset of components. Moreover, I assume for simplicity that the distribution of random variables is bivariate. An extension to multivariate distributions is tediously straightforward.

Let us consider *two* random and discrete variables, $H$ and $K$ with a list or record of events $h_1, h_2, …, h_i, …,$ and $k_1, k_2, …, k_j, …,$ respectively. The bivariate distribution is described by the values of probability for all *possible* pairs $(h_j, k_j)$ of random variables of the record $H$ and $K$.

Table 3. The bivariate distribution of discrete random variables $(h_j, k_j)$.

| ∴ | $h_1$ | $h_2$ | … | $h_i$ | … | $\Sigma_i$ |
|---|---|---|---|---|---|---|
| $k_1$ | $\mathcal{P}(h_1, k_1)$ | $\mathcal{P}(h_2, k_1)$ | … | $\mathcal{P}(h_i, k_1)$ | … | $\mathcal{P}_k(k_1)$ |
| $k_2$ | $\mathcal{P}(h_1, k_2)$ | $\mathcal{P}(h_2, k_2)$ | … | $\mathcal{P}(h_i, k_2)$ | … | $\mathcal{P}_k(k_2)$ |
| … | … | … | … | … | … | … |
| $k_j$ | $\mathcal{P}(h_1, k_j)$ | $\mathcal{P}(h_2, k_j)$ | … | $\mathcal{P}(h_i, k_j)$ | … | $\mathcal{P}_k(k_j)$ |
| … | … | … | … | … | … | … |

In Table 3, I have shown bivariate discrete distribution function of which the middle section contains the probabilities $\mathcal{P}(h_i, k_j)$ for all possible pairs $(h_j, k_j)$ with normalization condition $\sum_{i,j} \mathcal{P}(h_i, k_j) \equiv 1$. Probabilities $\mathcal{P}_h(h_i)$ and $\mathcal{P}_k(k_j)$ are called marginal probabilities. If all values of $\mathcal{P}(h_i, k_j)$ are known, in principle it is at least possible to calculate all marginal probabilities $\mathcal{P}_h(h_i) = \sum_j \mathcal{P}(h_i, k_j)$ and $\mathcal{P}_k(k_j) = \sum_i \mathcal{P}(h_i, k_j)$. However, the statement that if marginal probabilities $\mathcal{P}_h(h_i)$ and $\mathcal{P}_k(k_j)$ are known it is possible to recover $\mathcal{P}(h_i, k_j)$, is usually not true. It would only be true if marginals are independent random variables, $\mathcal{P}(h_i, k_j) = \mathcal{P}_h(h_i) \cdot \mathcal{P}_k(k_j)$, which generally is not the case for different meteorological variables.

For the observant reader, the above described marginals may appear to be Federov's "weather cases". This is indeed the case. The above scheme is the simplest two-dimensional discrete random variables case. Even in this case, though, the question about the distribution function of marginals arises. There are many different ways of presenting statistical distribution data and one of them is occurrence frequency distribution, or simply frequency distribution. I have already discussed wind event size distributions.

Before further discussion of marginals and their use in meteorology, one basic question must be addressed. Do marginal constructs lead to probability distributions? The answer to this question for normally distributed functions was provided by A. Sklar in 1959.[65] Let $\mathcal{P}$ be a joint distribution function with marginals $\mathcal{P}_1, \ldots, \mathcal{P}_d$. Then, there exists a copula $\mathcal{C}: [0,1]^d \to [0,1]$ such that for all $x_1, \ldots, x_d \in \overline{\mathbb{R}} = [-\infty, +\infty]$,

$$\mathcal{P}(x_1, \ldots, x_d) = \mathcal{C}(\mathcal{P}_1(x_1), \ldots, \mathcal{P}_d(x_d)). \quad (8)$$

One of the interesting properties of copulas is that if marginals are continuous, then $\mathcal{C}$ is unique.[66] Although construction of copulas for normally distributed variables is simple, the question of similar constructions for Lévy flights was not addressed by Sklar's theorem. There are a great number of processes described by Lévy flights including the above described wind events of the Antarctic. Because the power-law have a non-integrable singularity at 0, see eq.(6) one cannot construct copula in an ordinary way, and a copula for the tail integral $\overline{\Pi}$ was proposed.[67] In d-dimensional spectrally, positive Lévy process with marginal tail integrals $\overline{\Pi}_1, \ldots, \overline{\Pi}_d$, there exists a Lévy copula $\hat{\mathcal{C}}: [0, +\infty]^d \to [0, +\infty]$ such that for all $x_1, \ldots, x_d \in [0, +\infty]$,

$$\overline{\Pi}(x_1, \ldots, x_d) = \hat{\mathcal{C}}(\overline{\Pi}_1(x_1), \ldots, \overline{\Pi}_d(x_d)). \quad (9)$$

If the marginal tail integrals are continuous on $[0, +\infty]$, then the Lévy copula $\hat{\mathcal{C}}$ is unique since the copula induces a measure of probability on the Borel sets. The above theorem provides an important link between the positive Lévy process and marginal description of weather events. It tell us, that regardless of the underlying probability distribution it is possible to construct from distribution its marginals, the copula which exists and in a continuous case is unique.

The most poignant frequency distributions are log-log rank-order differential frequency distributions and cumulative frequency distributions. Returning to Table 3, it can be noted that all probabilities depicted in this table are not in any way sorted. One can sort the elements of this table in many different ways. For example, probability $\mathcal{P}(h_i, k_1)$ can be rank-ordered by ordering the elements according to increasing size. The biggest $\mathcal{P}(h_i, k_1)$ has the rank of $r_1 = 1$, the second biggest has a rank of $r_2 = 2, \ldots$, and the smallest has a rank of $r_n = n$, where $n = max(i)$. Since $\mathcal{P}(h_i, k_1)$ is not continuous, one cannot ignore the possibility of ties. However, in the case of reasonable statistical representation of distribution function by probabilities $\mathcal{P}(h_i, k_1)$, the possibility of ties between the ranks $r_i$ and $r_{i+1}$ is negligible and unimportant. Even if ties happen, their influence upon power-law curve fitting is diminutive. In the case of continuous distribution function, the possibility of ties can be ignored. Obviously in a rank-ordered sequence of $n$ events, the probability for the largest value is $1/n$, for events that are larger than the second-largest event, it is $2/n$, and so forth.

One can use the density function $p(x)$ (Eq. (2)) to calculate cumulative distribution function

$$\mathbb{P}(X \geq x) = \int_x^\infty p(y)dy = \frac{\varsigma(\alpha, x_{min})}{1-\alpha} x^{1-\alpha} \quad (10)$$

Let us assume, that $R$ is a sufficiently large sample drawn from the power-law distribution $p(\cdot | \alpha, x_{min})$, with $|R| = N$. According to the Law of Large Numbers,

$$\mathbb{P}(X \geq x) \approx \frac{|\{y \in R | y \geq x\}|}{N} \quad (11)$$

where $r_x = |\{y \in R | y \geq x\}|$ is a rank within R. Therefore, rank can be calculated and the power-law dependence is expressed by

$$\log x \approx (1-\alpha) \log r_x - \log \frac{N}{1-\alpha} \varsigma(\alpha, x_{min}). \quad (12)$$

In double logarithmic scale, power-law dependence is approximately the linear function of the rank $r_x$.

In his original work, Federov considered 12 meteorological parameters. For practical computational reasons, the considered number of parameters was shortened. Later on Ferdynus[68] used four meteorological variables, 4-tuple (temperature, cloudiness, precipitation, wind speed) = (T, N, R, W) to account for structure of weather states and their seasonality. Certain digits were attributed to the 4-tuple (T, N, R, W) and (T{1,…,9}, N{1,2,3}, R{0,1}, W{0,…,8}) was used to classify different weather types. The weather class defined as a temperature unbiased 3-tuple (N, R, W), was also studied.

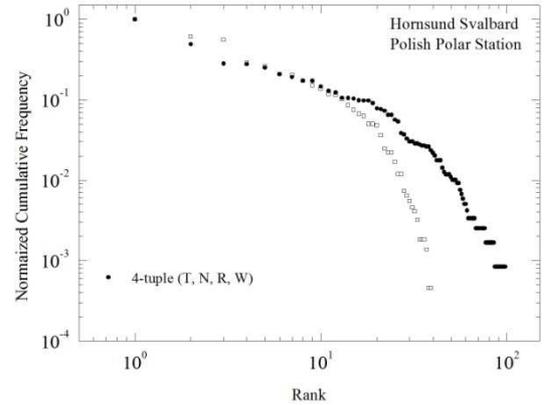

Figure 14. Cumulative distribution occurrence frequency of the 4-tuple (T, N, R, W) at the Hornsund Spitsbergen Island.

In Figure 14, cumulative distribution occurrence frequency of the 4-tuple (T, N, R, W) is depicted. It may be noticed that for high ranked 4-tuples, the dependence is almost linear. This information suggests that probability distribution of marginals is power-law like. However, as I have mentioned before, for accurate presentation, weather events such as 4-tuple (T, N, R,

W) must be binned to have constant logarithmic width, $b = \log(x_{i+1}) - \log(x_i)$. This binning procedure ensures a smooth appearance of event cumulative frequency distribution with little noise in the tail (fat tail). It also enables one to use a straightforward ordinary least squares method to find scaling parameters.

Thus, for the case of 4-tuple (T, N, R, W) at Hornsund weather station, the scaling parameter $\alpha = 0.92 \pm 0.03$ was calculated. This result combined with the preceding results, although obtained in a slightly different way, shows that the weather in Hornsund in Svalbard Island is self-organized criticality. It also experimentally proves the above mentioned Tankov-Kallsen theorem for positive Lévy process.

This power-law weather behaviour entirely confirms our previous calculations. Regardless of the time during the year, there is no *typical, average or mean* weather in the Hornsund. Moreover, because of the obtained scaling parameter $\alpha < 2$, the mean value of the considered weather types *cannot* simply be calculated. Consequently, the mean value of any variable in the 4-tuple (T, N, R, W) *cannot be calculated*. The variables – near surface air temperature, cloudiness, precipitation (rain and snow) and near surface wind – also describe self-organized phenomena. Consequently, one cannot attempt to calculate, for example, *mean* wind speed at a given location.

In the above presented analysis of the 4-tuple at Hornsund weather station, a natural time frame of 24h was used. There is quite a large fuzzy area in the definitions of the introduce variables. The end results are a poor approximation of reality. But, the Lévy copula $\hat{C}$ is unique and alpha-stable. Even if I have shortened the time frame to calculate 4-tuple, its copula $\hat{C}$ will have a similar attractor as it does not depend on a time frame but on asymptotic behavior of the copula.

Researchers, particularly meteorologists, are accustom to using all sorts of averages when calculating. The reader may be confused by the unrestrained use of the words: mean, average, and typical. I have shown above, that because the wind in the Antarctic as well as in similar settings is self-organized criticality, mean wind speed cannot be calculated or used. A similar comment is also applicable to temperature changes which are not Gaussian. There is a need to change a paradigm in meteorology by addressing fundamental epistemological questions. The main epistemological challenge to meteorology is to remove from its vocabulary locutions.[69]

## 8. Summary and Outlook

In this work, I have presented a summary of the analysis of near surface wind and temperature in the Antarctic. The presented analysis shows in an unequivocal way that near surface winds in the Antarctic are self-organized criticality phenomenon. The wind of the Antarctic is a preliminary air mixing agent. If this agent (driving force) is self-organized criticality, all reaming meteorological variables, as for example near surface temperature, snow transport, rate of snow melting, *etc*. must also be self-organized criticality phenomena.

The simple power-law describes the cumulative distribution function of wind event size, duration, and quiescent wind event measured at different locations as well as in mesoscale, in the Antarctic. The nature of these near surface winds is katabatic (drainage flow). The periods of stasis are also self-organized criticality. The observed self-organized criticality of katabatic winds suggests that atmosphere over the Antarctic continent is organized into many interacting katabatic air cells forming the Polar Cell – circumpolar near surface wind regime. The motion of these katabatic cells is driven by local rather than continental forces, predominantly by a downslope buoyancy forcing. Each air cell contributes to katabatic air cell at a certain threshold. Hence, the air over the Antarctic Plateau piles up on ice sheet formed gullies and valleys while relaxing its excess along contoured orography of coastal fringes or glacier slopes. At a given time, near surface air over the Antarctic continent contains a great number
of volatile air cells dissociated by close to equilibrium relaxed air cells, or the opposite.

Volatile and relaxed air cells are transient. These cells are formed much faster than the slow time-scale of driving force resulting from incoming solar radiation and the long-wave radiative loss to space. This process means the air system evolves through a sequence of states that are infinitesimally close to equilibrium in a cycle: driving, katabatic event, and relaxation. Transition between volatile air cells occurs through katabatic wind events which restore and redistribute driving field forces. The katabatic wind events, duration, and quiescent wind event have no preferred scale, and their sizes and durations follow power-law distributions. The numerical values of the exponents depend on the geographical location measured by the slope of the Antarctic dome. The wind field over the Antarctic is likely to be of a critical state. These are the essential factors of the mean field behaviour of near surface katabatic winds in the Antarctic or more generally over arbitrary ice dome on Earth or on other planets. Therefore, a fractional Fokker-Planck equation for the probability distribution of air particles whose motion is governed by a non-Gaussian Lévy-stable noise, would be a suitable theoretical approach in the description of near surface winds over the Antarctic continent.

## Acknowledgements

I wish to thank Professor Leszek Kułak from the Department of Theoretical Physics and Quantum Informatics at the Technical University of Gdańsk, Poland for assistance with neural network simulations. I also wish to thank him for the never ending discussions about entangled issues of the first principles of quantum mechanics. I am grateful to student Adrian Piasecki (M.Sc.) for his handling of weather data. The author is also grateful to Dr. Jacek Ferdynus form the Chair of Meteorology at Gdynia Maritime University for providing Hornsund weather data used in this work.

The author appreciates the support of the Antarctic Meteorological Research Center and the Automatic Weather Station Program for surface meteorological data, NSF grant numbers ANT-0636873, ARC-0713483, ANT-0838834, and/or ANT-0944018.

Finally, I wish to thank the editors of Monthly Weather Review for finding that "the field does not find your analysis compelling" and especially to Dr. Tom Hamill (NOAA Earth System Research Laboratory) and Dr. David M. Schultz (University of Manchester) who willingly participated in censorship, corruption (conflict of interest) of a review process and science.